# Entropic descriptor of a complex behaviour

R. Piasecki[1,*], A. Plastino[2,*]

[1] *Faculty of Chemistry, University of Opole, Oleska 48, 45-052 Opole, Poland*
[2] *La Plata Physics Institute (IFLP), National University La Plata and Argentina's National Research Council (CCT-CONICET), C.C. 727, 1900 La Plata, Argentina*

A B S T R A C T

We propose a new type of entropic descriptor that is able to quantify the statistical complexity (a measure of complex behaviour) by taking simultaneously into account the average departures of a system's entropy $S$ from both its maximum possible value $S_{max}$ and its minimum possible value $S_{min}$. When these two departures are similar to each other, the statistical complexity is maximal. We apply the new concept to the variability, over a range of length scales, of spatial or grey-level pattern arrangements in simple models. The pertinent results confirm the fact that a highly non-trivial, length-scale dependence of the entropic descriptor makes it an adequate complexity-measure, able to distinguish between structurally distinct configurational macrostates with the same degree of disorder.

*PACS*: 05.90.+m; 89.75.Kd

*Keywords*: Entropic descriptors; Statistical complexity; Multiscale analysis

## Contents

## 1. Introduction

Recently, a so-called versatile entropic measure (VEM) has been proposed [1]. This VEM is used for the multiscale analysis of grey level inhomogeneity (GLI) and is intended as a natural completion of the binary entropic measure $S_\Delta$ for extended objects [2,3]. VEM is based only on a combinatorial approach and employs Boltzmann's entropy. By recourse to the sliding cell-sampling (SCS) approach a striking effect was detected. Multiple intersecting curves (MIC) of the measure were encountered for paired simulated patterns differing, for instance, in the grey contrasting of sub-domains which were similar in size or symmetry properties [1,4]. This fact indicates a non-trivial dependence of the GLI on the length scale and suggests that the measure includes some features that may be useful for a multiscale variability analysis of complex patterns. In a binary case [5,6], the entropic measure $S_\Delta$ of spatial inhomogeneity was generalized to the Tsallis' entropy [7]. One can demonstrate (Cf. Appendix B in Ref. [5]) that, under certain conditions, the rate $S_\Delta / k^2$ displays similarities (at large length scales $k$) to the Shiner-Davison-Landsberg (SDL) entropic measure of complexity [8] denoted here as $C_{SDL}$, around which an

---

illuminating and enlightening discussion was carried out in [9–13]. On the other hand, the minimum value, $S_{min}$, of an entropy $S$, quite relevant for extended objects, is not taken into account neither by $S_\Delta/k^2$ nor by $C_{SDL}$, which are different functions of $S$ and $S_{max}$.

We will here advance a method that could be regarded as the natural starting point for the development of a universal and also practical multiscale entropic descriptor for the grey level or spatial complexity of various types of patterns. The method can be adapted so that different entropies be employed and is applicable to a wide range of systems. The basic ingredient is given in Eq. (1) below. We will illustrate the properties of our entropic descriptor by using Boltzmann's and Tsallis' entropies in the case of a few very simple systems.

## 2. The entropic descriptor

Ascertaining the degree of unpredictability and randomness of a system is not automatically tantamount to adequately grasp the correlational structures that may be present, i.e., to be in a position to capture the relationship between the components of the physical system. These structures strongly influence, of course, the character of the probability distribution that is able to describe the physics one is interested in. Randomness, on the one hand, and structural correlations on the other one, are not totally independent aspects of *this* physics. Certainly, the opposite extremes of i) perfect order and ii) maximal randomness possess no structure to speak of. In between these two special instances a wide range of possible degrees of physical structure exists, degrees that should be reflected in the features of the underlying probability distribution. One would like that they be adequately captured by some functional of the pertinent probability distribution in the same fashion that Shannon's entropy captures randomness. A suitable candidate to this effect has come to be called the *statistical complexity* (see the helpful discussion of [14]).

The most common versions of statistical complexity [15–22] give it the form of a product. One multiplies the actual system's entropy times the distance from the associated probability distribution (PD) to the uniform PD, called the "disequilibrium". In most cases, the ensuing complexity measure is neither intensive nor extensive. Since several "distance-forms" can be concocted, many possibilities are open.

To avoid such multiplicity we propose here the following intensive general form for our entropic descriptor of complex behaviour (CB), that i) entirely bypasses the need for a disequilibrium and ii) vanishes for perfect order or complete randomness, namely,

$$C_\lambda = \frac{1}{\lambda} \frac{(S_{max} - S)(S - S_{min})}{(S_{max} - S_{min})}, \qquad (1)$$

where $S$, $S_{max}$ and $S_{min}$ refer, respectively, to the actual entropy, and to its maximum and minimum values for a given system, while $\lambda$ is a parameter related to the averaging-procedure to be employed. For a given pattern of size $L \times L$, the parameter $\lambda$ denotes, typically, the number of cells $k \times k$ pertaining to the specific partitioning procedure one has selected. This fact allows one to compare (for a given system or systems) the different descriptor values at different length scales $k \in \{1, 2,\ldots, L\}$. It is then reasonable to regard the parameter as a function of $k$, i.e., $\lambda \to \lambda(k)$, and also $C_\lambda \to C_\lambda(k)$. The descriptor becomes then a length-scale depending quantity. We may as well consider other types of system for which, instead of using a given *configurational* type of entropy per cell, one utilizes Shannon's or Tsallis's entropies. Then, making use of entropies per microstate, one could also compare systems with a different number $\lambda \to W$ of states. Depending on what specific comparison purposes we focused on, there are thus many possibilities of giving $\lambda$ a specific meaning.

Keeping in mind the general form of the entropic descriptor one can easily check that, by using the definitions $x = (S_{max} - S)/\lambda \equiv S_\Delta$ (Cf. [2,3]), $y = (S - S_{min})/\lambda$, and $x + y = (S_{max} - S_{min})/\lambda = \delta$, one can write

$$C_\lambda = \frac{xy}{(x+y)} = \left(x - \frac{x^2}{\delta}\right) = \left(y - \frac{y^2}{\delta}\right), \qquad (2)$$

where $x \in [0, \delta]$ and $0 \leq C_\lambda(x) \leq C_{\lambda, max}(x_0) = \delta/4$ at $x_0 = \delta/2$ (Cf. Fig. 1). In practice, the equivalent notation, $C_{\lambda, max}(S)|_{S=S_0} = (S_{max} - S_{min})/(4\lambda)$ and $S_0 = (S_{min} + S_{max})/2$ can be usefully employed.

The most CB, or in "pattern's language", the most complex arrangement at a given length scale emerges when the average *departures* of the actual entropy $S$ from the highest one $S_{max}$ and from the lowest reference entropy $S_{min}$ are similar to each of other, a kind of compromise between two opposite limiting configurations: the most homogeneous and the most inhomogeneous. For these relatively uncomplicated cases the descriptor $C_\lambda$ tends to it's the lowest values. Within a linear approximation the corresponding boundary expressions are given by

$$C_\lambda \cong \begin{cases} \dfrac{(S_{max} - S)}{\lambda} \equiv S_\Delta & \text{for} \quad x \to 0^+, \\ \dfrac{(S - S_{min})}{\lambda} \equiv \delta - S_\Delta & \text{for} \quad x \to \delta^-. \end{cases} \qquad (3)$$

We recognize that the upper formula has been used previously for our investigations of various types of degree of patterns-inhomogeneity [1−6], using $S_\Delta$ together with a microcanonical entropy, $S(k) = k_B \ln \Omega(k)$, where the Boltzmann constant $k_B = 1$ for convenience, $\Omega(k)$ being the number of (proper) configurational microstates. Such an approach allows for rewriting of Eq. (1) in another, rather enlightening form, namely,

$$C_\lambda = \frac{1}{\lambda} \frac{\ln(\Omega_{max}/\Omega)\ln(\Omega/\Omega_{min})}{\ln(\Omega_{max}/\Omega_{min})} \ . \qquad (4)$$

Now, in this representation $C_{\lambda, max}(\Omega)|_{\Omega=\Omega_0} = [\ln(\Omega_{max}/\Omega_{min})]/(4\lambda)$ for $\Omega_0 = (\Omega_{max}\Omega_{min})^{1/2}$.

When pixels are not treated as points, binary or greyscale patterns belong to an important class of the systems: finite-size objects (FSO). For such systems, the most inhomogeneous arrangement differs from that obtaining for point objects. For instance, for binary patterns at a given length scale $k > 1$, it is impossible to place all black pixels inside a single-cell of size $k \times k$ (in contrast to what happens with points). If we consider greyscale patterns within the context of the so-called pillar model [23] there arises another restriction entailing more complicated mathematics. Therefore, usually we consider $S_{min}(k) > 0$, a condition that is advantageous for the comparison of patterns at different length scales by recourse to our entropic descriptor $C_\lambda$. If we compare the normalized measures $C_{SDL}/C_{SDL, max} = 4\Delta(1 - \Delta)$, with $\Delta \equiv S/S_{max}$ [8] and $C_\lambda/C_{\lambda, max}$, the significance of the non-zero term $S_{min}$ become apparent. A toy-model example is found in the dashed lines of Fig. 4(b), and also in the B-macrostates of Tab. 1 (Appendix). In general, the entropy-based measures of generalized inhomogeneity (including also the measure $C_\lambda$ of CB) seem to provide more structural information than the two-point spatial correlation function alone, Cf. Fig. 5 in Ref. [23].

On the other hand, in the case of patterns composed of points or particle systems with particles approximated by points, we become less constrained with regards to the lowest value of configurational entropy. In particular, when an evolving $W$-state system is described with an appropriate probability distribution, e.g., $(p_1, p_2,..., p_W)$, the zero value of its entropy can be easily attributed to the $(0, 0,...,0, 1)$-instance. If just a single state can be occupied we have always $S_{min} = 0$. Thus, the normalized entropic descriptor $C_\lambda / C_{\lambda, max}$ reduces itself to the form $4\Delta(1 - \Delta)$ [8].

Generally, taking into account the properties of the our entropic descriptor, $C_\lambda$ can be placed within the second category of statistical complexity measures described in [8], [17−19], that covers measures which can be quite small for larger amounts of either order or disorder, with a maximum at some intermediate stage.

## 3. Examples

In order to examine the validity of our approach, we will apply it to a few simple systems with a small number $W$ of microstates. We employ here the microcanonical Tsallis entropy for three values of the concomitant non-extensivity index $q$: 0.6, 1, and 1.4. The intermediate value unity yields the extensive Shannon entropy-instance. According to Eq. (1) a convenient final $q$-form of the entropic descriptor can be cast as

$$C_{\lambda=W} = \frac{1}{W(q-1)} \frac{(W^{1-q} - \sum_{i=1}^{W} p^q)(1 - \sum_{i=1}^{W} p^q)}{(W^{1-q} - 1)} , \qquad (5)$$

where $S_{q, min} = 0$ is assumed. Let us begin by depicting the universal parabolic shape of the entropic descriptor $C_{\lambda=W}(x; \delta)$ of Eq. (2), using Tsallis' as the entropic-quantifier. In Fig. 1, instead of using a particular $\delta$-value we compute Eq. (2) as applied to a few instances of two and three states systems with $W = 2$ and 3.

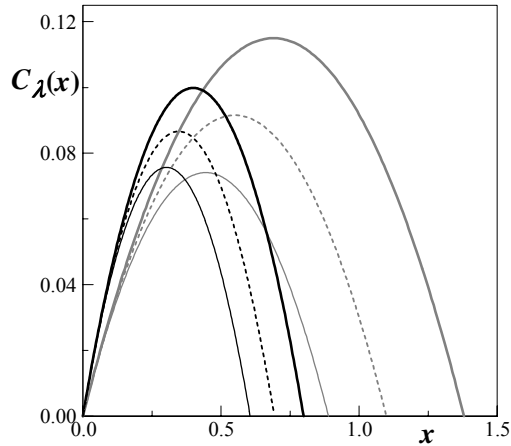

**Fig. 1.** Universal parabolic shape of the entropic descriptor $C_{\lambda=W}(x; \delta)$ represented as a function of $x = (S_{max} − S)/\lambda \equiv S_\Delta$, Cf. [2,3]. The bold, dashed, and thin lines for a system of two states ($W = 2$), and their grey counterparts for a system of three states ($W = 3$), correspond, respectively, to $q = 0.6$, 1 (equivalent to the Shannon case), and 1.4. One appreciates the fact that the theoretical maximum of complexity $C_{\lambda, max}(x_0) = \delta/4$ is expected always to be located at $x_0 = \delta/2$.

Next, we test again a two-state system ($W = 2$) but for any possible choice of occupation probabilities, $p$ and $1 - p$ (Cf. Fig. 2).

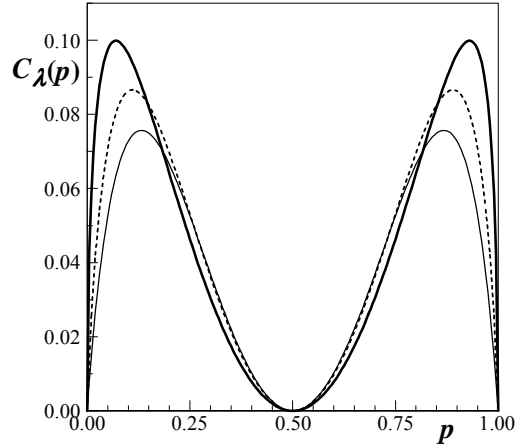

**Fig. 2.** The entropic descriptor $C_{\lambda=W}(p)$ for a system of two states ($W = 2$) occupied with probabilities $p$ and $1 - p$. The bold, dashed and thin lines correspond to the Tsallis index $q = 0.6$, 1 (equivalent to the Shannon case) and 1.4, respectively. In each case there are two symmetrically located probability values $p^*$ and $1 - p^*$ that yield the most complex behaviour, $p^*(q = 0.6) \cong 0.07$ with $C_{\lambda, max} \cong 0.10$, $p^*(q = 1) \cong 0.11$ with $C_{\lambda, max} \cong 0.087$, and $p^*(q = 1.4) \cong 0.13$ with $C_{\lambda, max} \cong 0.077$. For the uniform distribution (½, ½) the system's complexity vanishes with $S_q = S_{q, max}$ and the same happens in the case of the maximally non-uniform one [(0, 1) and (1, 0)], but now with $S_q = S_{q, min}$. Interestingly enough, for $p \in (p_a, p_b)$ or $(1 - p_b, 1 - p_a)$ with $p_a \cong 0.1465$ and $p_b \cong 0.2655$, the value of $C_\lambda(p;$ Shannon$)$ exceeds that of $C_\lambda(p;$ Tsallis$)$.

Remarkably enough, the behaviour of the entropic descriptor $C_\lambda(p)$ in Fig. 2 is similar to that of the dynamical complexity defined as a variant of the predictability (see the tent map example in Fig. 2(a) of [24], where it is plotted as a function of the skewness parameter). As expected, in all cases $C_\lambda(p)$ is small both near the equiprobable distribution and also for cases of large non-uniformity. At the two symmetrically located probability-values, $p^*(q)$ and $1 - p^*(q)$, a $C_\lambda$-maximum is reached indicative of very CB. However, our intuition fails in predicting for which specific entropic-quantifier will the complexity be greater. Firstly, there are symmetrical $p$-"points" at which the domination of $C_\lambda(p;$ Tsallis, $q=0.6)$ over $C_\lambda(p;$ Tsallis, $q=1.4)$ is reversed (and vice versa). Secondly, there exist $p$-intervals for which $C_\lambda(p;$ Shannon$)$ is greater than its two $C_\lambda(p;$ Tsallis, $q)$ counterparts.

In a similar vein we pass now to consider a three-state system for any possible value of the occupation probabilities, $p_1$, $1 - p_1$, and $1 - p_1 - p_2$, Cf. Figs. 2(a) and (b) for non-extensivity index $q$-values 0.6 and 1.4, respectively. As expected, even if the $C_\lambda(p_1; p_2)$-behaviour displays a more complicated geometry, it remains qualitatively similar to that exhibited in the preceding example. Instead of isolated, single-probability values, for each case here there exist three separate contour-lines (not shown) built out of appropriate pairs $(p_1^*, p_2^*)$, that signal the existence of maximal CB.

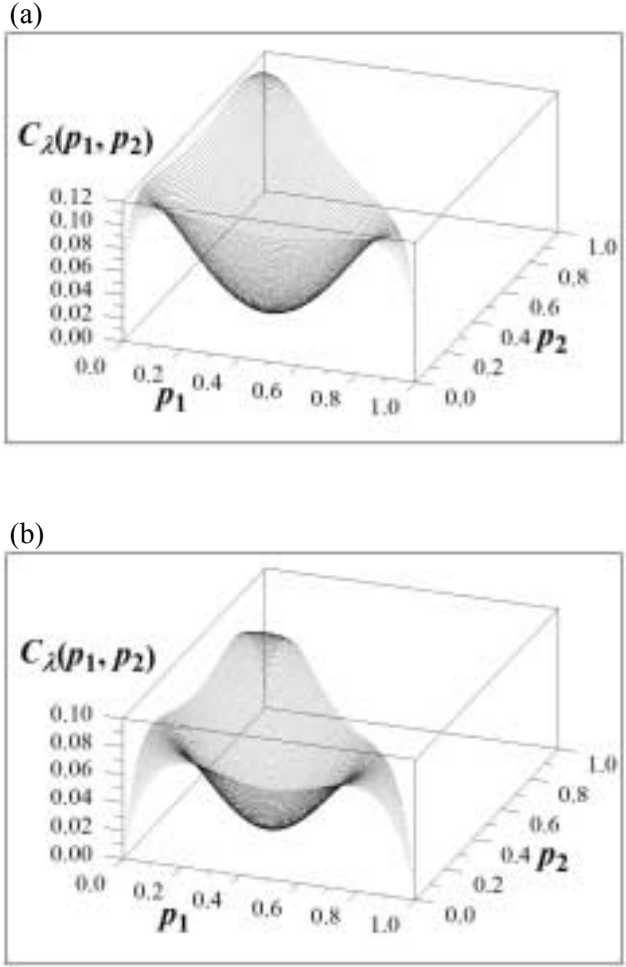

**Fig. 3.** The entropic descriptor $C_{\lambda=W}(p_1, p_2)$ for a system with three states ($W = 3$) occupied with probabilities $p_1$, $p_2$ and $1 - p_1 - p_2$. The surface corresponds to a Tsallis' index (a) $q = 0.6$ and (b) $q = 1.4$. In each instance there are three places at which the contour-lines built out of appropriate pairs $(p_1^*, p_2^*)$ signal maximal CB of the same "degree". Notice that $C_{\lambda, max}(q = 0.6) \cong 0.115 > C_{\lambda, max}(q = 1.4) \cong 0.074$. For the Shannon case (not depicted) we have intermediate value of $C_{\lambda, max} \cong 0.092$. For the uniform distribution (⅓, ⅓, ⅓) the system's complexity equals to zero because of $S_q = S_{q, max}$. Also, as expected from the symmetry reasons, the same behaviour there is for three possible maximally non-uniform distributions [(0, 0, 1), (0, 1, 0), and (1, 0, 0)] because of $S_q = S_{q, min}$.

Now, the CB-spatial aspects for comparable systems are illustrated at common length scale. For each of our illustrative examples, all possible system's macrostates are clustered into representative classes, as listed in Tab. 1 of the Appendix. Certain macrostates are also represented by configurations of black pixels, which are placed on a very small lattice partitioned (in a standard way) at the fixed length scale $k = 2$. Fig. 4 depicts the normalized entropic descriptor $C_\lambda / C_{\lambda, max}$ as a function of the configurational entropy $S$ for a few model-systems of different number of finite-size objects. A characteristic feature of spatially complex systems becomes apparent. Using the notation of Tab. 1, the largest complexity for macrostates C#6 and A#4 is maximal since the corresponding entropies are located exactly at $S_0 = (S_{min} + S_{max})/2$, see the remark following the Eq. (2). Also the rather unusual situation encountered for the B#3 and B#4 macrostates, both with the same entropy and the largest complexity, involves an entropic value $S = 4.1589$ which is the closest possible one to the theoretical $S_0 = 4.0740$.

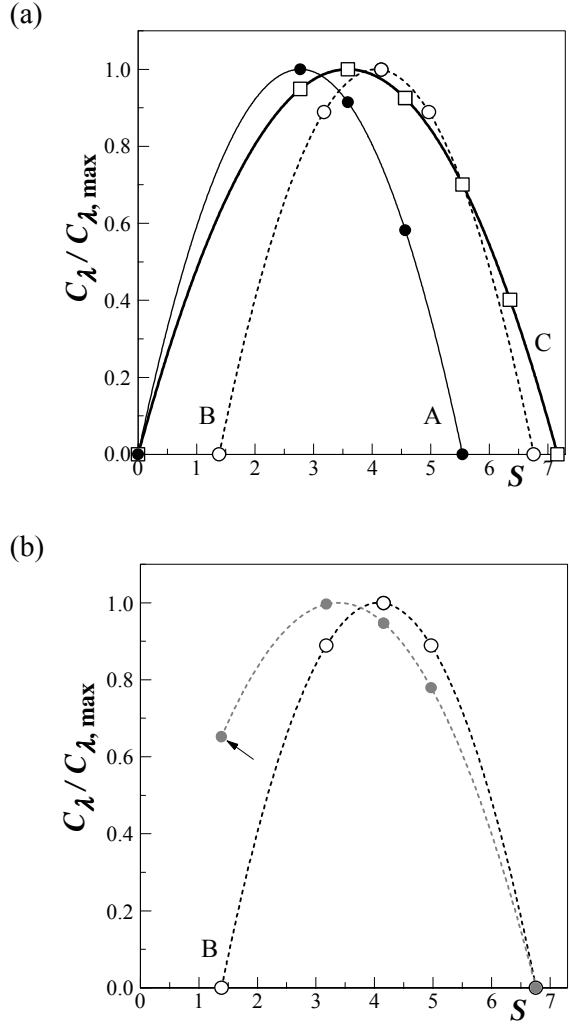

**Fig. 4.** The normalized entropic descriptor $C_\lambda(S)/C_{\lambda,\,max}$ as a function of configurational entropy $S$ for a $4 \times 4$-lattice partitioned into $\lambda = 4$ non-overlapping cells, at length scale $k = 2$, for a given number $N$ of black pixels. The symbols refer to representative classes of configurational macrostates. (a) The C-bold, B-dashed and A-thin lines correspond to $N = 8$, 7 and 4, respectively. Some of the identical-entropy macrostates compared for cases C and A exhibit different CB, while in case B we observe certain degenerations. (b) for case B ($N = 7$) the two measures, $C_\lambda(S)/C_{\lambda,\,max}$ [black] and $C_{SDL}/C_{SDL,\,max}$ [grey] dashed lines, are compared. The arrow indicates the maximally ordered macrostate (B#6) to which a non-vanishing complexity is attributed by the SDL-measure, in contrast to the more reasonable zero-value of $C_\lambda$. Other details can be extracted from Tab. 1 in the Appendix.

However, with the help of an SCS calculation (overlapping cells), as exemplified in the Appendix, one finds that $C_\lambda(\text{B\#3; SCS}) = 0.2940 < C_\lambda(\text{B\#4; SCS}) = 0.2947$. It is also worth noticing that $S(\text{C\#6}) = S(\text{A\#3})$ but $C_\lambda(\text{C\#6}) > C_\lambda(\text{A\#3})$ and, analogously, $S(\text{C\#7}) = S(\text{A\#4})$, although $C_\lambda(\text{C\#7}) > C_\lambda(\text{A\#4})$. This entails that, at a fixed scale, two comparable systems with exactly the same disorder can still differ in their CB, as quantified by $C_\lambda/C_{\lambda,\,max}$. Additionally, for the maximally ordered macrostate (B#6) indicated by an arrow in Fig. 4, the SDL-measure attributes to it a non-vanishing complexity, in contrast to the zero-value complexity given by $C_\lambda$. The next example illustrates an even more striking feature: for a given system containing macrostates of quite similar disorder-degree, the entropic descriptor $C_\lambda$ can distinguish between their respective structural complexities.

Thus, we pass to focus attention on examples displaying spatially more complex configurational macrostates, but at different length scales. In Fig. 5(a) we consider a specific configuration of black pixels placed on a larger lattice but using still the SCS approach at all possible length scales $k$. Notice that this approach is less sensitive to fractal properties as compared to the standard partition into non-overlapping cells. We consider next the interesting, structurally deterministic Sierpinski carpet (DSC), see the upper inset in Fig. 5(a). Afterwards, we modify its structure in two ways so as to try to detect possible changes in the sensitivity of the entropic descriptor. Both the DSC's pseudo-random counterpart (RSC) (with conserved structural black parts) and the random pattern (RPA) of $1 \times 1$ objects that ensues after some memory-erasing of the initial structure, are depicted in the middle and bottom insets, respectively. We are still able to detect, by looking at the "bold" curve, traces of typical behaviour like positions of peaks, minima, and even shape-self-similarity, characteristic of the DSC.

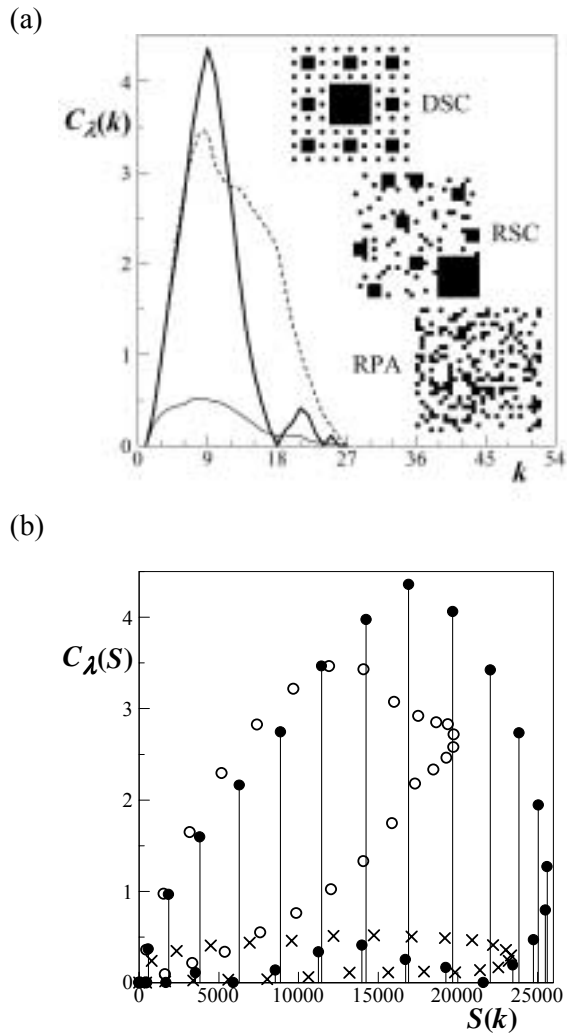

**Fig. 5.** We display the entropic descriptor $C_\lambda$ in various ways making use of the microcanonical entropy, $S(k) = \ln \Omega(k)$, that does not distinguishes mutually reversed binary patterns (white $\leftrightarrow$ black) [2,3]. (a) as a function of the length scale $k$ for the inverted deterministic Sierpinski carpet (DSC) of size $27 \times 27$ (in pixels), [bold line], its pseudo-random counterpart (RSC) with conserved sizes of square objects, [dashed line], and for a corresponding random pattern (RPA) of $1 \times 1$ objects, [thin line]; (b) as a function of the microcanonical DSC-entropy [drop line], RSC [open circles] and RPA [crosses]. Increasing randomness for the RSC and RPA cases significantly reduces $C_\lambda(k)$, ie., the degree of CB around the first peak. In turn, our descriptor $C_\lambda(S)$ clearly discriminates, for a variety of systems, the distinct structural complexity of different macrostates, although they are characterized by quite similar disorder-degrees.

Despite the simplicity of these patterns one can indeed notice a non-trivial length scale dependence of the entropic descriptor, This is confirmed, additionally, by the effect of multiply intersecting curves (MIC) observed also in the context a of grey-level inhomogeneity in [1]. Unexpectedly, for $k > 12$ the spatial complexity of the RSC pattern is greater than that of the initial DSC-one. This is not always true for the RPA case. Moreover, the entropic descriptor $C_\lambda(S)$, even for rather simple systems with two different macrostates of nearly identical disorder-degree (described at different length scales), can still discriminate between their spatial complexities, Cf. Fig. 5(b). On the other hand, the corresponding characteristic shapes $C_\lambda(S)$ seem to be diffused for the simplest RPA-case, opposite to what happens for the more structurally complex DSC-case.

(a)

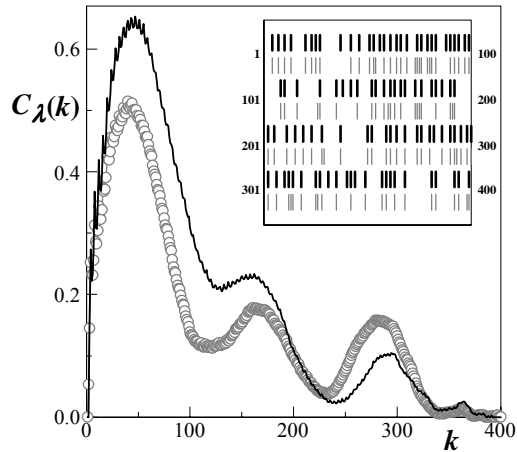

(b)

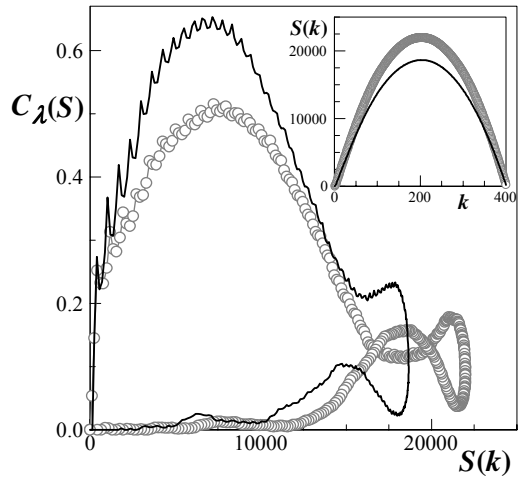

**Fig. 6.** Our entropic descriptor is plotted in various ways: (a) as a function of the length scale $k$ for a 1D lattice-gas random configuration (called RND) of 100 particles, see the thin vertical grey lines (on $L = 400$ possible locations) and [grey circles], and also the thick black lines for the RND-counterpart, called RLG, that includes a "nearest neighbour exclusion" rule; (b) as a function of the (exactly calculated) microcanonical RLG-entropy [thick black line] and RND-one [grey circles] (the SCS approach is employed). In spite of the weak nature of the RLG-correlations caused by the "no-two-particles adjacent"-rule, acting as a kind of repulsive interaction, the degree of complex behaviour $C_\lambda(k)$ does get increased when we use the first half of our length scales. Also, the behaviour of the two $C_\lambda(S)$ becomes intricate. Moreover, for each of the two 1D patterns of the graph there exist scales $k_i \leq k_j$ at which the involved entropies are nearly identical (see the inset and text) but the corresponding complexities $C_\lambda(S)$, instead, are significantly different for RND-case (notice the "crossings" and see the text).

Fig. 6 refers to yet another simple test of our entropic CB-descriptor. In Ref. [20] the authors underline the fact that a useful complexity measure should be the sensitive to the role of a system's correlations. Therefore, it would be interesting to compare the length scale behaviour of $C_\lambda(k)$ for a system of interacting hard-core particles [25], whose authors show that stripe formation may result from a purely repulsive isotropic short-range pair potential with two characteristic length scales. Here we propose the much simpler but effective test of the 1D lattice gas, by employing two-pattern configurations, in the spirit of the interesting work reported in [26], devoted to 2D case, but without numerical results.

Our computer program generated a random-configuration (RND) by randomly "tossing" 100 particles (denoted by thin vertical grey lines in Fig. 6(a)) onto a line with $L = 400$ locations, subject only to the constraint that two of them could not simultaneously fall onto the same location. A second configuration, called here RLG (repulsive lattice gas), is generated in exactly the same manner except for the addition of the additional rule that two particles (denoted by thick vertical black lines in Fig. 6(a)) cannot fall onto adjacent locations. For the two cases the same random "seed" was employed. The alluded constraints induce correlations that simulate a kind of repulsive interaction. Such correlations are, in the case of the first half of the length scales we use, sufficiently strong as to increase the degree of a spatially CB [see RLG (thick black line)] as compared to the RND instance (grey circles) in Fig. 6(a). It should be stressed that the exact values of configurational entropies $S(k; RLG)$, as represented by the thick black line in the inset of Fig. 6(b), as well as the quantities $S_{max}(k; RLG)$ and $S_{min}(k; RLG)$, are computed using adequate combinatorial formulae for the number $[\Omega(k; RLG)]$ of realizations of a given macrostate given the above mentioned constraint. Such a formula for a 1D case can be cast in the simple form

$$\Omega(k; RLG) = \prod_{i=1}^{\kappa} \binom{k - n_i + 1}{n_i} \leq \prod_{i=1}^{\kappa} \binom{k}{n_i}, \qquad (6)$$

where $\kappa(k) = L - k + 1$ denotes allowed positions of the sliding cell of size $1 \times k$ and $n_i \leq (k + 1)/2$ describes the number of particles occupying the $i$th sampling-cell. Equality in Eq. (6) happens only for the smallest scale $k = 1$, when a "nearest neighbour exclusion" rule does not apply. Thus, the inequality $S(k; RLG) < S(k; RND)$ holds at every length scale $1 < k \leq L$. For example, $S(k = L; RLG) = 188.338$ while $S(k = L; RND) = 221.856$ (Cf. also the inset in Fig. 6(b)). This confirms the general conclusion of [26], based on the argument that an RND-configuration is typical in the class comprising a larger number of configurations. There are not that many available in the RLG-case.

Particularly intricate becomes, in this example, the behaviour of the two $C_\lambda(S)$ in Fig. 6(b). One can find scales $k_i \leq k_j$ at which the involved entropies are nearly identical but the corresponding complexities $C_\lambda(S)$, instead, are significantly different. For instance, $S(k = 156; RLG) = 17623.117$ and $S(k = 250; RLG) = 17623.036$, while the corresponding complexities are $C_\lambda(k = 156; RLG) = 0.2249$ and $C_\lambda(k = 250; RLG) = 0.0282$. Similarly, one can find $S(k = 195; RND) = 21948.835$ and $S(k = 212; RND) = 21948.569$, while the respective $C_\lambda(k = 195; RND) = 0.1153$ and $C_\lambda(k = 212; RND) = 0.0659$. Thus, once again, essentially distinct configurations (RND) and (RLG) with nearly the same amount of disorder can be clearly distinguished at distinct length scales by our descriptor because their respective spatial complexities are different. In addition, for the RND case, and due to the fact that the third peak of $C_\lambda(k; RND)$ is of a strength comparable to that of the second one, the corresponding $C_\lambda(S; RND)$ in Fig. 6(b) undergoes multiple self-intersections. There are three intervals of length scales at which a curious interplay between $C_\lambda(S; RND)$ and $S(k; RND)$ can be observed.

As follows from the definition of configurational entropy [2,3], see also the right-hand side of inequality in Eq. (6), the entropy-value does not change under the replacement of a "black phase" (with concentration $\varphi$) by a "white phase" (with concentration $1 - \varphi$), and vice versa. Thus, for the inverted patterns of all the binary images above [except for those patterns related to RLG-configurations, whose specific correlational properties are not conserved under a white ↔ black interchange of pixels], the same $C_\lambda(k)$ curves are obtained, a particular lack of sensitivity that can be overcome when we deal with grey-level images. This why in a final example we test our entropic descriptor, still using the SCS approach, for the structurally interesting greyscale pattern adapted from [27]. An initial pattern of size $151 \times 151$ (in pixels) is depicted in Fig. 7(a) (Ts1-inset). The pattern's morphology is dictated by specific ordering mechanisms at work in confined diblock copolymers. Here however, we are interested in its non-trivial structural grey level periodicity with its further modifications.

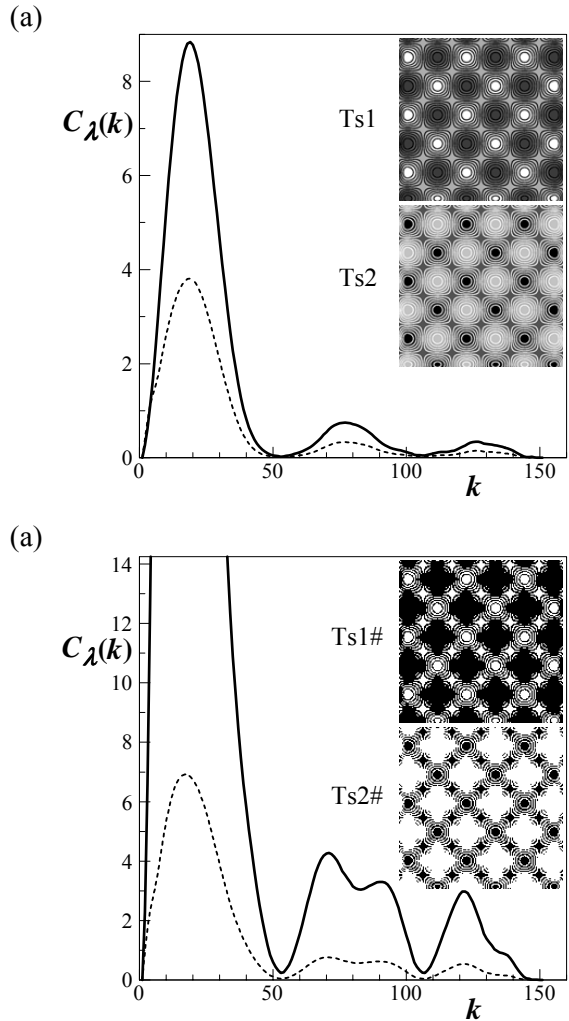

**Fig. 7.** We display the entropic descriptor $C_\lambda(k)$ as a function of the length scale $k$ making use of the microcanonical entropy, $S_{gr}(k) = \ln \Omega_{gr}(k)$, as proposed for greyscale patterns in Ref. [1]. (a) for the grey scale pattern Ts1 of size $151 \times 151$ (in pixels) adapted from [27] [bold line], and its converted ($i \to 255 - i$) grey scale counterpart Ts2 [dashed line]. (b) Same as (a), but for the two associated binarized patterns (see the text) Ts1# and Ts2#. In the case of the Ts1#-pattern, the maximum of about 70.56 is not shown so as to better visualizing the dashed line. Within the SCS approach one can observe traces of periodicity (two deep minima whose separation is the same for all curves) and shape-self-similarity around the first peak. The descriptor $C_\lambda(k)$ distinguishes between the complexities of the Ts1 and Ts2 patterns linked via a kind of symmetry-operation, in grey scales, and does so for the two associated binarized patterns, Ts1# and Ts2#. Additionally, the binarization procedure reveals two neighbouring peaks at intermediate values of $k$.

First, the initial pattern is converted by a simple symmetry operation [$(i \rightarrow 255 - i)$ in grey levels] into its greyscale counterpart, see the Ts2-inset in Fig. 7(a). Then, the simplified "binarization" procedure ($i \rightarrow j$) is applied to those patterns: if grey level index $i < 128$, then it becomes a black one with $j = 0$, otherwise grey pixels becomes white with $j = 255$. This procedure gives two associated binarized patterns (although differently encoded vis-a-vis the zero-one standard binary matrix), see the insets Ts1# and Ts2# in Fig. 7(b). For the corresponding patterns in Figs. 7(a) and (b) we observe two pairs of curves (similar in shape around the position of the first peak): Ts1 (Ts1#) [bold lines] and Ts2 (Ts2#) [dashed lines]. This is so because of the method employed for the construction of the $S_{gr}(k)$ entropy for grey-level patterns. The approach utilizes all possible order-dependent partitions of grey level values over $k^2$ positions inside each cell. In mathematics this is sometimes referred to as a *weak composition* [28]. Thus, the entropic descriptor $C_\lambda(k)$ becomes also dependent on the total sum of grey level values. At a given length scale, in order to make a *quantitative* comparison of the CB of grey-scale patterns differing in their total sums, we should calculate the entropic descriptor per "grey level". In our case such a procedure results in an increasing of the values for the bold curves, but it leaves unchanged the sequence of the considered pairs of curves. On the other hand, the "binarization" of grey-scale images, that leads to two colour images encoded in grey-scale fashion, becomes quite useful in revealing some details of the CB at large scales. For instance, in Fig. 7(b) one can observe two neighbouring peaks at an intermediate range of $k$-values which are not detected in Fig. 7(a).

Summing up, we underline the fact that our entropic CB-descriptor allows for clearly distinguishing non-random variations in the pattern's structure from its random counterparts at different length scales, as seen, for instance, in Figs. 5 and 6. However, since at this stage only spatial CB was dealt with, we have to emphasize the role played by the kind of partitioning used. This point is intimately linked to the behaviour of the entropic descriptor itself as size varies [29] and thus deserves further investigation.

## 4. Conclusions

We have advanced an entropic descriptor $C_\lambda$ in Eq. (1) that generalizes the SDL measure of statistical complexity [8]. The basic properties of our descriptor have been illustrated by recourse to a rather variegated sample of simple systems. Making use of Tsallis' entropy for two- and three-state systems, the expected behaviour and characteristic features of a structural complexity measure was clearly observed in Figs. 2 and 3. Using Boltzmann's entropy for representative classes of simple configurational macrostates, the expected diversity in localization on the universally shaped curve of entropic descriptors, when plotted against the entropy, was indeed reproduced in Fig. 4 and Tab. 1.

In Fig. 5, that deals with i) the structurally deterministic Sierpinski carpet, ii) its pseudo-random counterpart, iii) and a fully random case, we studied in the case of standard examples just how the degree of structurally CB, as measured by $C_\lambda(k)$, diminishes for many length scales $k$. The more complicated behaviour of $C_\lambda(k)$ and $C_\lambda(S)$ is illustrated by Fig. 6, where the role of a system's correlations was in evidence. The physically different configurations of the 1D lattice gas, a) the random (RND) and b) the partially random (RLG), were clearly discriminated by their respective spatial complexities, that display a particularly intricate behaviour.

Finally, we employed the recently introduced grey-level entropy to the case of i) a suitably adapted pattern and ii) its transformed (into a grey-level fashion) counterpart, which were further subjected to the specific "binarization" procedure. In Fig. 7 we saw how the descriptor $C_\lambda(k)$ properly distinguishes among the distinct complexities in all instances, uncovering also a non-trivial, structural grey level periodicity. For the binarized patterns the descriptor additionally detects certain peaks, invisible for the initial greyscale patterns. This augurs well for the possibility of an enhancement of the entropic descriptor sensitivity at larger scales. A quite complicate dependence of the entropic descriptor on the length scale was detected in all relevant figures.

The main conclusion to be drawn from these examples can be summarized as follows: structurally distinct configurations with nearly the same amount of disorder-degree can be distinguished at different length scales by our entropic CB-descriptor because their respective complexities are different.

**Appendix**

**Table 1.** Collection of representative classes of macrostates* and their associated i) entropies, ii) entropic descriptor $C_\lambda$, and iii) relative form $C_\lambda/C_{\lambda,\,max}$ [and $C_{SDL}/C_{SDL,\,max}$ in case (B)] for a toy model with: (A) $N = 4$, (B) $N = 7$. and (C) $N = 8$. The black pixels are placed on a $4 \times 4$ lattice partitioned into $\lambda = 4$ (not overlapping) cells at length scale $k=2$. The maximal values of the relative complexities are given in bold-faced format. The last columns include also results of a $C_\lambda(SCS)$-calculation (using the sliding cell-sampling approach) for the specific representative configurations given below.

| Case | Macr.# | Config. | $S_{min}$ | $S$ | $S_{max}$ | $C_\lambda$ | $C_\lambda/C_{\lambda,\,max}$ | $C_{SDL}/C_{SDL,\,max}$ | $C_\lambda(SCS)$ |
|---|---|---|---|---|---|---|---|---|---|
| A | 1 | 1 1 1 1 | | 5.5452 | 5.5452 | 0.0 | 0.0 | | |
| A | 2 | 0 1 1 2 | | 4.5643 | | 0.2018 | 0.5823 | | |
| A | 3 | 0 0 2 2 | | 3.5835 | | 0.3169 | 0.9144 | | |
| A | 4 | 0 0 1 3 | | 2.7726 | | 0.3466 | **1.0** | | 0.2759 |
| A | 5 | 0 0 0 4 | 0.0 | 0.0 | | 0.0 | 0.0 | | |
| B | 1 | 1 2 2 2 | | 6.7616 | 6.7616 | 0.0 | 0.0 | 0.0 | |
| B | 2 | 0 2 2 3 | | 4.9698 | | 0.2986 | 0.8889 | 0.7791 | |
| B | 3 | 0 1 3 3 | | 4.1589 | | 0.3356 | **0.9989** | 0.9470 | 0.2940 |
| B | 4 | 1 1 1 4 | | 4.1589 | | 0.3356 | **0.9989** | 0.9470 | 0.2947 |
| B | 5 | 0 1 2 4 | | 3.1781 | | 0.2986 | 0.8889 | **0.9964** | |
| B | 6 | 0 0 3 4 | 1.3863 | 1.3863 | | 0.0 | 0.0 | 0.6520 | |
| C | 1 | 2 2 2 2 | | 7.1670 | 7.1670 | 0.0 | 0.0 | | |
| C | 2 | 1 2 2 3 | | 6.3561 | | 0.1798 | 0.4014 | | |
| C | 3 | 1 1 3 3 | | 5.5452 | | 0.3137 | 0.7003 | | |
| C | 4 | 0 2 3 3 | | 4.5643 | | 0.4144 | 0.9251 | | |
| C | 5 | 1 1 2 4 | | 4.5643 | | 0.4144 | 0.9251 | | |
| C | 6 | 0 2 2 4 | | 3.5835 | | 0.4479 | **1.0** | | 0.3386 |
| C | 7 | 0 1 3 4 | | 2.7726 | | 0.4250 | 0.9553 | | |
| C | 8 | 0 0 4 4 | 0.0 | 0.0 | | 0.0 | 0.0 | | |

* e.g., for A#4 the notation 0013 denotes representative macrostate realized by

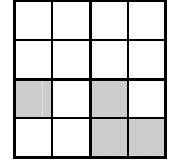

$\binom{4}{0}\binom{4}{0}\binom{4}{1}\binom{4}{3} = 16$ configurational microstates, one of them being $\Rightarrow$

This macrostate exhibits the highest value of $C_\lambda = 0.3466$ for case (A).
For the above specific representative configuration one can create the corresponding macrostate (using SCS-tenets), i.e., 000111123, having 6144 realizations. Thus, the value of the entropic descriptor will be $C_\lambda(SCS) = 0.2759$. Now, for B#3, i.e., for the 0133 representative macrostate one obtains

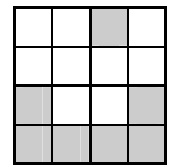

$\binom{4}{0}\binom{4}{1}\binom{4}{3}\binom{4}{3} = 64$ configurational microstates, one of them being $\Rightarrow$

This macrostate and the one displayed below exhibit the highest possible value $C_\lambda$ = 0.3356 for case (B), while for the corresponding macrostate 011101323 (SCS used again) one finds $C_\lambda$(SCS) = 0.2940. The associated degenerate B#4, i.e., the 1114 macrostate is realized by

$$\binom{4}{1}\binom{4}{1}\binom{4}{1}\binom{4}{4} = 64 \text{ configurational microstates, one of them being} \Rightarrow$$

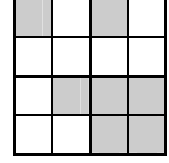

This macrostate exhibits, as previously, a highest value of $C_\lambda$ = 0.3356. Using the SCS approach, i.e., for the corresponding macrostate 111122134, we obtain $C_\lambda$(SCS) = 0.2947, which differs from the previous one. This means that certain degenerations can be removed with SCS-help. In turn, the C#6 case, i.e., the 0224 macrostate, is realized by

$$\binom{4}{0}\binom{4}{2}\binom{4}{2}\binom{4}{4} = 36 \text{ configurational microstates, one of them being} \Rightarrow$$

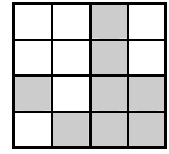

This macrostate exhibits the highest possible value $C_\lambda$ = 0.4479 for case (C) of this toy model with $1 \leq N \leq 16$ at length-scale $k$ = 2. For the corresponding SCS-macrostate, i.e., 022123234 one finds $C_\lambda$(SCS) = 0.3386.